\documentclass{new_tlp}
\usepackage{booktabs}
\usepackage[T1]{fontenc}
\newcommand\djordjename{\DJ or\dj e Markovi\'{c}\xspace} 
\usepackage{ifthen}
\usepackage{url}
\usepackage{amssymb}
\usepackage{amsmath}
\usepackage{import}
\usepackage{xparse}
\usepackage{amsmath}
\renewcommand*\ttdefault{cmvtt}
\usepackage[usenames,dvipsnames]{xcolor}
\usepackage{xspace}
\usepackage{datatool}
\usepackage{glossaries}

\newcommand{\m}[1]{\ensuremath{#1}\xspace}
\newcommand{\trval}[1]{\m{\mbox{\bf #1}}}

	\newcommand{\limplies}{\m{\Rightarrow}}
	\newcommand{\limpl}{\limplies}
	\newcommand{\lequiv}{\m{\Leftrightarrow}}

	\newcommand{\lrule}{\m{\leftarrow}}
	\newcommand{\cause}{\m{\stackrel{c}{\lrule}}}
	
	\newcommand{\ltrue}{\trval{t}}
	\newcommand{\lfalse}{\trval{f}}
	\newcommand{\lunkn}{\trval{u}}
	
	\newcommand{\voc}{\m{\Sigma}}

	\newcommand{\struct}{\m{I}}

	\newcommand{\theory}{\m{\mathcal{T}}}

	\newcommand{\D}{\m{\Delta}}

	\newcommand{\set}{\m{S}}
	\NewDocumentCommand\inter{g+g}{%
	  \IfNoValueTF{#1}
	    {\struct}
	    {\m{#1^{#2}}}}

	\newcommand{\xxx}{\m{\overline{x}}}

	\newcommand{\ttt}{\m{\overline{t}}}

	\newcommand{\nat}{\m{\mathbb{N}}}
	\newcommand{\Nat}{\nat}
	\renewcommand{\int}{\m{\mathbb{Z}}}

	\newcommand{\leqp}{{\m{\,\leq_p\,}}}

	\newcommand{\kleeneval}{\m{Kl}}

	\NewDocumentCommand\subs{g+g}{%
	  \IfNoValueTF{#1}
	    {\m{/}}
	    {\m{#1/ #2}}}

	\newcommand{\logicname}[1]{\text{\sc #1}\xspace}

	\newcommand{\idp}{\logicname{IDP}}

	\newcommand{\idpthree}{\logicname{IDP$^3$}}
	\newcommand{\idpdraw}{\logicname{ID$^{P}_{Draw}$}}

	\newcommand{\minisatid}{\logicname{MiniSAT(ID)}}

	\newcommand{\foid}{\logicname{FO(\ensuremath{ID})}}
	
	\newcommand{\fo}{\FO}

\newcommand{\ouracronym}[3]{%
	\newacronym{#1}{#2}{#3}
	\expandafter\newcommand\csname #1\endcsname{\gls{#1}\xspace}%
}
	\ouracronym{FO}{FO}{first-order logic}
	\ouracronym{MX}{MX}{Model Expansion}
	\ouracronym{MO}{MO}{Model Optimization}
	\ouracronym{ASP}{ASP}{Answer Set Programming}
	\ouracronym{TP}{TP}{Theorem Proving}
	\ouracronym{LP}{LP}{Logic Programming}
	\ouracronym{CP}{CP}{Constraint Programming}
	\ouracronym{KR}{KR}{Knowledge Representation}
	\ouracronym{CSP}{CSP}{Constraint Satisfaction Problem}
	\ouracronym{SMT}{SMT}{SAT Modulo Theories}
	\ouracronym{KBS}{KBS}{knowledge-base system}
	\ouracronym{NNF}{NNF}{Negation Normal Form}
	\ouracronym{FNNF}{FNNF}{Flat Negation Normal Form}
	\ouracronym{DefNNF}{DefNNF}{Definition Negation Normal Form}
	\ouracronym{CDCL}{CDCL}{Conflict-Driven Clause-Learning}
	\ouracronym{LCG}{LCG}{Lazy Clause Generation}

	\makeatletter
	\def\ifenv#1{
	\def\@tempa{#1}%
	\def\@ttempa{#1*}%
	\ifx\@tempa\@currenvir
	\expandafter\@firstoftwo
	\else
	\expandafter\@secondoftwo
	\fi
	}
	\makeatother

	\newcommand{\ddrule}[4]{\ensuremath{#1 \leftarrow #2 & \{#3\} & #4}}
	\newcommand{\drule}[2]{\ensuremath{#1 & \leftarrow & #2}}

	\newcommand{\darule}[4]{\ensuremath{#1 \leftarrow #2 & \{#3\} & #4}}
	\newcommand{\arule}[2]{\ensuremath{#1 \, &\leftarrow \, #2}}

	\newenvironment{ldef}{\renewcommand\arraystretch{1.25}\left\{\begin{array}{l@{ \,}l@{\,}l}}{\end{array}\right\}}
	\newenvironment{ltheo}{\[\begin{array}{l}}{\end{array}\]\ignorespacesafterend}

	\newcommand{\LNDRule}[2]{
	\ifenv{array}
	{\drule{#1}{#2}}
	{ \ifenv{align}
		{\arule{#1}{#2}}
		{\ifenv{align*}
		{\arule{#1}{#2}}
		{ERROR: using LDRule in unsupported environment: \@currenvir}
		}
	}
	}

	\newcommand{\LDRule}[4]{
	\ifenv{array}
	{\ddrule{#1}{#2}{#3}{#4}}
	{ \ifenv{align}
		{\darule{#1}{#2}{#3}{#4}}
		{\ifenv{align*}
		{\darule{#1}{#2}{#3}{#4}}
		{ERROR: using LDRule in unsupported environment: \@currenvir}
		}
	}
	}

	\NewDocumentCommand\LRule{m+g+g+g}{%
		\IfNoValueTF{#2}%
		{#1.&}{%
		\IfNoValueTF{#3}
		{\LNDRule{#1}{#2.}}
		{\LDRule{#1}{#2.}{#3}{#4}}%
		}
	}

	\NewDocumentCommand\CLRule{m+g}{%
	\ifenv{array}
	{\cdrule{#1}{#2}}
	{ \ifenv{align}
		{\carule{#1}{#2}}
		{\ifenv{align*}
			{\carule{#1}{#2}}
			{ERROR: using CLRule in unsupported environment: \@currenvir}
		}
	}
	}

	\NewDocumentCommand\carule{m+g}{%
		\IfNoValueTF{#2}
			{\ensuremath{#1.}}
			{\ensuremath{#1 \, &\cause \, #2}}}
	\NewDocumentCommand\cdrule{m+g}{%
		\IfNoValueTF{#2}
			{\ensuremath{#1.}}
			{\ensuremath{#1 & \cause & #2}}}

	\newcommand{\algrule}[4]{
	\hbox{{#1}:}& 
	\quad #2 ~\longrightarrow~ #3 
	\hbox{~ if } #4\\
	}

	\newcommand{\AlgoRule}[4]{
	\ifenv{array}
	{\algrule{#1}{#2}{#3}{#4}}
		{ERROR: using AlgoRule in unsupported environment: \@currenvir}
	}

\newcommand{\commentstyle}{\color{Gray}}
	\usepackage[final]{listings}

	\lstdefinelanguage{idp}{
		morekeywords=[1]{namespace,vocabulary,theory,structure,procedure,term,set,formula, spec, specification},
		morekeywords=[2]{include,using,type,isa,contains,partial,extern,LFD,GFD,constructed,from,constraint,func,pred,supertype,of,subtype,define},
		morekeywords=[3]{int,float,char,string,nat},
		morekeywords=[4]{if,then,else,for,end},
		morecomment=[s]{/*}{*/},	
		morecomment=[l]{//}
	}
	\newcommand{\ignore}[1]{}

	\newboolean{nocomments}
	\setboolean{nocomments}{false}

	\newboolean{commentmargin}
	\setboolean{commentmargin}{true}

	\newcommand{\namedcomment}[3]{
		\ifthenelse{\boolean{nocomments}}
		{} %IF no comments, write nothing
		{ %Otherwise
			\ifthenelse{\boolean{commentmargin}}
				{ {\color{#3} \marginpar{\color{#3}\sc #2}#1}  } %Name in margin
				{  {\color{#3} {\sc #2}: #1}  } %Name not in margin
		}
	}
	\newcommand{\mnamedcomment}[3]{\ifthenelse{\boolean{nocomments}}{}{{\marginpar{ \color{#3}{\sc #2}:#1}}}}

	\usepackage{soul}

\newcommand{\keyword}[2]{%
	\expandafter\newcommand\csname #1\endcsname{#2\xspace}%
	\expandafter\newcommand\csname #1s\endcsname{#2s\xspace}%
	\expandafter\newcommand\csname #1ness\endcsname{#2ness\xspace}%
}

\newcommand{\domainof}[1]{\m{D^{#1}}}

\newtheorem{thm}{Theorem}[section]
\newtheorem{theorem}[thm]{Theorem}

\newtheorem{corollary}[thm]{Corollary}
\newtheorem{proposition}[thm]{Proposition}

\newtheorem{defn}[thm]{Definition}
\newtheorem{definition}[thm]{Definition}

\newtheorem{example}[thm]{Example}

\newtheorem{ex*}{Example}

\renewcommand{\time}{\m{Time}}
\newcommand{\timed}{dynamic\xspace}
\newcommand{\start}{\m{\mathcal{I}}}
\newcommand{\next}{\m{\mathcal{S}}}
\newcommand{\proj}[1]{\m{#1_{curr}}}
\newcommand{\pron}[1]{\m{#1_{next}}}
\newcommand{\Vs}[1]{\m{#1_{s}}}
\newcommand{\Vss}[1]{\m{#1_{ss}}}
\newcommand{\Vbs}[1]{\m{#1_{bs}}}
\newcommand{\pacman}{Pac-Man\xspace}
\newcommand{\vocp}{\m{\voc^p}}
\newcommand{\Position}{\m{Pos}}

\renewcommand\struct{\m{J}}

\newcommand{\inittheo}{\m{\theory_0}}
\newcommand{\transtheo}{\m{\theory_t}}
\newcommand{\transl}[1]{\m{te(#1)}}

\newcommand{\pss}[2]{\m{\pi^{ss}_{#2}(#1)}}
\newcommand{\pbs}[2]{\m{\pi^{bs}_{#2}(#1)}}
\newcommand{\p}[2]{\m{\pi_{#2}(#1)}}

\newcommand\extension[2]{\m{#1\hspace{-3pt}::\hspace{-3pt}#2}}

\newcommand\bcmdtab{\noindent\bgroup\tabcolsep=0pt%
  \begin{tabular}{@{}p{10pc}@{}p{20pc}@{}}}
\newcommand\ecmdtab{\end{tabular}\egroup}

  \title[Simulating Dynamic Systems Using LTC]
        {Simulating Dynamic Systems Using Linear Time Calculus Theories}

  \author[B. Bogaerts et al.]
         {Bart Bogaerts, Joachim Jansen, Maurice Bruynooghe, \and Broes De Cat, Joost Vennekens, and Marc Denecker \\
         Department of Computer Science, KU Leuven\\
         \email{\{firstname.lastname\}@cs.kuleuven.be}}

\pubyear{2014}
\pagerange{\pageref{firstpage}--\pageref{lastpage}}

\begin{document}

\label{firstpage}

\maketitle

%  \setboolean{nocomments}{true} %Remove comment to remove comments
% % 

 \begin{abstract}
	Dynamic systems play a central role in fields such as
planning, verification, and databases.  Fragmented throughout these
fields, we find a multitude of languages to formally specify dynamic
systems and a multitude of systems to reason on such
specifications. Often, such systems are bound to one specific language
and one specific inference task. 
It is troublesome that performing several inference tasks on the same
knowledge requires translations of your specification to other languages. 
In this paper we study whether it is possible to perform
a broad set of well-studied inference tasks on one specification.
More concretely, we extend \idpthree with several inferences from fields
concerned with dynamic specifications.
% 
% The idea of the knowledge base paradigm is that knowledge should be 
% represented once in a truly declarative
% logic and different reasoning tasks are performed by applying
% different inference methods on the same knowledge. In this paper, we
% show that the knowledge base paradigm is applicable in the context of 
% dynamic systems by extending \idpthree so that it can handle a
% wide range of reasoning tasks relevant for dynamic systems.

  \end{abstract}


% \todo{theorem proving zonder streepje}
% \todo{LTC-theory met streepje}
% \todo{\pacman met streepje}
% \todo{\pacman game zonder streepje}
% \todo{check aantal occurrences of ``will'' en haal er wat weg}
% \joachim{references naar secties enzo zijn geschreven met een hoofdletter (bvb: Section 3.4), moet dat met een hoofdletter?} \bart{JA}

% \todo{zoek stukken dubbele tekst, sommige zaken worden te vaak uitgelegd}

% \tableofcontents

% \todo{Some glueing sentences between the definitions would
%   help to increase readability.}
% 
% \todo{Some evaluation on real examples could help to see if this approach
%   scales in practice. But is exceeds the scope of the paper and can be done in
%   future work.}

\paragraph{Note} This is an updated version of a paper published in 2014 in Theory and Practice of Logic Programming. The original paper had a small mistake in Theorem 4.4, which is rectified in the current version. Essentially, only one of two implications holds. This has no impact on other claims in the paper. 
We are grateful to \djordjename for pointing out this mistake. 

\section{Introduction}\label{sec:intro}
  
Traditionally, systems that reason on declarative specifications take input in a specific language 
and perform one specific inference on this input. 
As argued in \citeN{iclp/DeneckerV08} and \citeN{ASPOCP/Denecker12}, 
it is often useful to perform several different inference tasks on the same knowledge base; the authors called 
this idea the Knowledge Base System (KBS) paradigm. 
In this paper, we evaluate the usefulness of the KBS
paradigm in the well-studied domain
 of dynamic action languages. 
We identify many interesting inference tasks in this domain and we 
show that for one concrete action language, the Linear Time Calculus (LTC) introduced in this paper, each of these tasks can be performed. 
As a result, we can use the same specification,
%\joachim{Wat bedoel je precies met ``code base''? Is dit hetzelfde als problem specification? Als dit hetzelfde is gebruik je best dezelfde term hier denk ik.} 
an LTC-theory, for
performing a wide range of tasks, whereas traditional software
development uses different specifications 
for different tasks. We illustrate
this with different tasks related to development of a Pac-Man game.

% We will not start from scratch, but our general approach will reduce inference tasks for LTC theories to existing inference methods. 
We do not start from scratch, our general approach reduces inference tasks for LTC-theories to existing inference methods.
We do this in the context of the \idpthree system  \cite{WarrenBook/DeCatBBD14},
% a knowledge-based programming environment: 
a KBS that allows
manipulation of logical objects (theories, structures, terms,
queries,\dots) through an imperative layer and hence allows users to
glue the different inference methods together to construct useful
software  \cite{IDPKBS}. For \idpthree, the imperative layer is the  Lua
scripting language \cite{SPE/IerusalimschyFC96}.

Many action calculi have been developed, including the Situation Calculus \cite{McCHay69,book/Reiter01}, Event Calculus \cite{ngc/KowalskiS86}, and the Planning Domain Definition Language (PDDL) \cite{ghallabGCPSSSSD98} 
and many inference methods exist for these calculi.
\emph{Progression inference} aims at finding the successor of a given state, and plays a central role in database applications \cite{journals/ai/LinR97,DBLP:journals/corr/KowalskiS13}. %  given an action theory .
\emph{Simulation} uses progression to simulate the execution of a system based on a formal specification.
%The \emph{planning} community is concerned with finding a sequence of actions that achieve a goal. 
\emph{Planning} aims at finding a sequence of actions that achieve a goal.
Other inference tasks are  finding and/or proving \emph{invariants} of
a dynamic system, and \emph{verifying} complex temporal statements such as LTL or CTL expressions.

Even though many important inference methods---including
theorem proving \cite{Fitting96}, (optimal) model expansion
\cite{lash08/WittocxMD08}, querying \cite{Vardi86} and debugging
\cite{kbse/ShlyakhterSJST03}---are already supported by \idpthree, several 
inferences in a dynamic context, such as progression, simulation, and proving invariants, are not yet supported.
% \old{
To overcome this limitation, we show that all of the above inferences can be implemented in a subclass of theories, the so-called Linear Time Calculus (LTC) theories.
% \joachim{Eerste vermelding van \foid, hier zou een uitleg bij moeten: In this paper we show that \idpthree's formal specification language, \foid, is sufficiently expressive to describe a wide range of dynamic systems. To this end, we introduce a subclass of \foid-theories, the so-called Linear Time Calculus (LTC) theories.}

%
From a dynamic (time-dependent) LTC-theory, we derive two simpler static  theories---an
initial theory and a transition theory---and we show that progression 
on the LTC-theory can be performed by model expansion on the
transition theory and that simulation can be achieved by repeated progression. Proving invariants is achieved by induction: by proving it for the
initial theory and proving that the property is preserved by the transition theory.
Finally we discuss how one could handle more complex dynamic properties.

% 
% 
% \ignore{
% \joost{stuk over software development eventueel in de eerste paragraaf mergen?}
% We situate all of the above inferences in the context of
% logic-based software development and use a \pacman game as running
% example.  In standard software development, specification, \todo{broes vond dat specification te veel op abstraction leek}
% implementation and the abstraction used for verification are three
% distinct descriptions of the software. Here we argue that the same
% code base, the LTC-theory, can be used for all these tasks. 
% This is much less error-prone than
% the traditional approach.  We show that the LTC-theory of our \pacman
% game is a specification that can be used to play the game by an
% interactive simulation that reads the user input to progress the state
% of the game.  The same theory is also the basis for proving
% invariants. Finally, playing the game automatically can be seen as a planning problem
% solved by model expansion on the LTC-theory.\broes{let op met volgordes}
% }

The main contributions of this paper are threefold: 
i) we illustrate the practical advantages of the KBS paradigm in the context of dynamic systems,
ii) we implement methods to perform progression and simulation and to prove invariants on the same theory, and 
iii) we study the relation between various declarative problem-solving domains concerned with dynamic systems and identify which inferences are studied in which domains.

The paper is organised as follows. 
In Section \ref{sec:prelims} we recall some preliminaries. 
Next, in Section \ref{sec:LTC}, we introduce the class of structures and theories of interest:
structures describing an evolution of a state over time and theories that essentially contain only local information.
Afterwards, in Section \ref{sec:inferences}, we provide an overview of the inferences applicable to these theories and 
in Section \ref{sec:comparison} we compare with other systems.
We conclude in Section \ref{sec:conclusion}. 
Omitted proofs can be found in Appendix B.

\section{Preliminaries}\label{sec:prelims}
  % \bart{check wat er hiervan nodig is en wat niet}
We assume familiarity with basic concepts of \FO. 
% Vocabularies, formulas, and terms are defined as usual. %; we use $\varphi\limplies \psi$ as a shorthand for $\lnot\varphi\lor \psi$. 
%  For the sake of simplicity, we will not allow to reuse the same variable within quantifications (i.e., expressions of the form 
%  $\forall x: \exists x: P(x)$ will be excluded in this text.
% A \voc-structures interprets all symbols (including variable symbols) in \voc; % \domainof{\I} denotes the domain of \I and
If \struct is a structure over \voc, and $\sigma$ a symbol in vocabulary \voc, $\sigma^\struct$  denotes
the interpretation of $\sigma $ in \struct. 
% If \voc is a vocabulary, a \voc-structure $\I$ consists of a domain \domainof{\I} and assigns a value 
% $\sigma^\I$ to all symbols $\sigma$ in \voc (including variable symbols).
% We use $I[\subs{v}{\sigma}]$ for the structure $\J$ that equals \I, except on $\sigma$: $\sigma^\J=v$.
% \emph{Domain atoms} are atoms of the form $P(\ddd)$ where the $d_i$ are domain elements.
If $\varphi$ is a formula and $t_1$ and $t_2$ are terms, we use $\varphi[\subs{t_2}{t_1}]$ for the formula obtained from $\varphi$ by replacing all occurrences of $t_1$ by $t_2$.
We use a many-typed logic and write $P(t_1,\dots,t_n)$ and $f(t_1,\dots,t_n): t'$ for the predicate $P$ typed $t_1,\dots,t_n$ respectively the function $f$ with input arguments of type $t_1,\dots,t_n$ and output argument typed $t'$. %\mbroes{Well-typedness?}

\foid extends \fo with (inductive) definitions: sets of rules of the form 
$\forall \xxx: P(\ttt)\lrule \varphi,$ (or $\forall \xxx: f(\ttt)=t'\lrule \varphi$)
where $\varphi$ is an \fo formula and the free variables of $\varphi$ and $P(\ttt)$ are among the $\xxx$. We call $P(\ttt)$ (respectively $f(\ttt)=t'$) the head of the rule and $\varphi$ the body.
The connective $\lrule$ is the \emph{definitional implication}, which should not be confused with the material implication $\limpl$. Thus, the expression $\forall \xxx: P(\ttt)\lrule \varphi$ is \emph{not} a shorthand for $\forall \xxx: P(\ttt)\vee \neg \varphi$. 
Instead, its meaning is given by the
well-founded semantics (for functions, semantics of the graph predicate is considered, i.e., as if the rule were $Graph_f(\ttt,t)\lrule\varphi$); 
this semantics correctly formalises all kinds of definitions 
that typically occur in mathematical texts \cite{tocl/DeneckerT08,DeneckerV14}.

To simplify the presentation, we assume, without loss of generality, that an \foid theory
consists of a set of FO sentences and a single definition 
\cite{tocl/DeneckerT08,jelia/MarienGD04}; our 
implementation does not impose this restriction.

% \todo{functie definities}

\section{Linear Time Calculus}\label{sec:LTC}
  
We define linear-time vocabularies and structures. 
Next, we define the progression
inference  and analyse when progression can be performed without
keeping an explicit history 
 (when a theory satisfies the Markov property~\cite{markov1906}). 
Finally, we define the Linear Time Calculus, and show that LTC-theories  satisfy the Markov property. 

% we will describe the syntactical restrictions on an \foid theory for it to be well-formed in the Linear Time Calculus (LTC). LTC-theories will describe how a dynamic system evolves in time similar to the  Situation Calculus \cite{McCHay69}: by describing state transitions. The big difference with Situation Calculus is that LTC, as the name suggests, reasons about linear time, while Situation Calculus is branching time logic. This means that a model of an LTC-theory corresponds to one simulation of a dynamic system, while models in Situation Calculus represent the entire tree of possible executions.

%%%%%%%%%%%%%%%%%%%%%%%%%%%%%%%%%%%%%%%%%%%%%%%%
%%%%%%%%%%%%%%%%%%%%%%%%%%%%%%%%%%%%%%%%%%%%%%%%
%LTC Vocabularies
%%%%%%%%%%%%%%%%%%%%%%%%%%%%%%%%%%%%%%%%%%%%%%%%
%%%%%%%%%%%%%%%%%%%%%%%%%%%%%%%%%%%%%%%%%%%%%%%%

\subsection{Linear-Time Vocabularies and Structures}
% \maurice{TODO: choose ``sort'' or ``type''?, type is IDP terminologie} \bart{maar sort lijkt meer standaard in Logica...}

\begin{definition}[Linear-time vocabulary]\label{def:ltcvoc}
% A \emph{linear-time vocabulary} is a many-typed first-order vocabulary \voc satisfying the following restrictions:
A \emph{linear-time vocabulary} is a many-typed first-order vocabulary
\voc such that:
 \begin{itemize}
  \item \voc has a type $\time$ (always interpreted as \nat), a constant $\start$ of type $\time$ (interpreted as $0$) and a function $\next(\time):\time$ (interpreted as the successor function),
  \item All other symbols in \voc have at most one argument of type \time,
  \item Apart from \start and \next, the output argument of
  functions is not of type  \time.
 \end{itemize}
We partition symbols in \voc in three categories: \time, \start and \next are LTC-symbols,
 symbols without a \time argument are \emph{static} symbols, and all other symbols are \emph{dynamic} symbols.
For ease of notation, we will assume that the \time always occurs last in dynamic symbols.
\end{definition}
In the rest of this paper we  assume that \voc is a linear-time vocabulary. 
% We restrict our attention to \voc-structures interpreting \time by the natural numbers, \start by the initial time ($0$), and \next by the successor function (mapping every $t$ to $t+1$).
Such a structure describes an evolution of a state over time, i.e., it represents a sequence of states.
Here, a state is a structure over a vocabulary derived from \voc by projecting
out \time.

\begin{definition}[Projected symbol]
If $\sigma(t_1,\dots, t_{n-1},\time)$ is a dynamic predicate
  symbol, then
  $\proj{\sigma}(t_1,\dots, t_{n-1})$  is its
  \emph{projected symbol}. Similarly, for a function symbol
  $\sigma(t_1,\dots, t_{n-1},\time):t$, its \emph{projected symbol} is $\proj{\sigma}(t_1,\dots,
  t_{n-1}):t$.
\end{definition}

\begin{definition}[Derived vocabularies]
The vocabularies derived from \voc are:
\begin{itemize}
	\item the \emph{static vocabulary} \Vs{\voc} consisting of all static symbols in $\voc$;
	\item the \emph{single-state vocabulary} \Vss{\voc} which extends
  the static vocabulary \Vs{\voc} with the symbol $\proj{\sigma}$ for each \timed symbol
  $\sigma $ in \voc.
% 	\item 
\end{itemize}
\end{definition}
Intuitively, a \Vss{\voc}-structure describes a single state, and $\proj{\sigma}$ describes the interpretation of $\sigma$ on that point in time. 

\begin{example}\label{ex:pacmanvoc}
A simplification of the \pacman game can be modelled with a linear-time vocabulary $\vocp$ consisting of types $\time$, $Agent$, $Square$, and $Dir$.
The vocabulary contains predicate symbols
  $Next(Square,Dir,Square)$, $Move(Agent,$ $ Dir, \time)$,
  $Pell(Square, \time)$, and $GameOver(\time)$. An atom $Next(s,d,s')$
  expresses that the square next to $s$ in direction $d$ is $s'$,
  $Move(a,d,t)$ that agent $a$ moves in direction $d$ at time $t$,
  $Pell(s,t)$ that square $s$ contains a pellet at time $t$, and
  $GameOver(t)$ that either all pellets are eaten or \pacman is dead at $t$.  The vocabulary
  also contains a constant $pacman$ of type $Agent$  and function
  symbols $\Position(Agent,\time):Square$ mapping each agent
  at every time-point to his position, and $StartPos(Agent):Square$
  mapping agents to their initial position.

\Vs{\vocp} and \Vss{\vocp} %and The derived vocabularies 
have the same types as \vocp.
\Vs{\vocp} contains the constant $pacman$, the predicate symbol $Next(Square,Dir,Square)$,
 and the function $StartPos(Agent):  Square$.
%.
\Vss{\vocp} is an extension of \Vs{\vocp} with predicate symbols $\proj{Move}(Agent, Dir)$, $\proj{Pell}(Square)$, and $\proj{GameOver}$  and the function symbol $\proj{\Position}( Agent):Square$.
\end{example}

%%%%%%%%%%%%%%%%%%%%%%%%%%%%%%%%%%%%%%%%%%%%%%%%
%%%%%%%%%%%%%%%%%%%%%%%%%%%%%%%%%%%%%%%%%%%%%%%%
%LTC STRUCTURES
%%%%%%%%%%%%%%%%%%%%%%%%%%%%%%%%%%%%%%%%%%%%%%%%
%%%%%%%%%%%%%%%%%%%%%%%%%%%%%%%%%%%%%%%%%%%%%%%%
% 
% Structures over the various vocabularies are very related. 
% A \voc-structure describes an evolution of a state over time. \todo{staat er dubbel in}
% A \Vss{\voc}-structure describes the state of the dynamic system at one timepoint, i.e., it describes a snapshot of the world. 
% A \Vbs{\voc}-structure describes the state transation between two successive time points.
% 
% We define some useful mappings between them.

 \begin{definition}[Projection of a structure]

Let $\Omega$ be the interpretation of a dynamic symbol $\sigma/n$ in a structure and $k\in\nat$.
   The set of tuples $(d_1,\dots,d_{n-1})$ such that
 $(d_1,\dots,d_{n-1},k)\in \Omega$ is the
 \emph{k-projection} of $\Omega$, denoted $\p{\Omega}{k}$.
%
%  \end{definition}
%
% \begin{definition}
% We define two projections on \voc-structures. 

The \emph{single-state
  projection} of a \voc-structure \struct on time $k\in\nat$, denoted  $\pss{\struct}{k}$, is the $\Vss{\voc}$-structure
interpreting static symbols $\sigma$ as $\sigma^\struct$ and 
dynamic symbols $\proj{\sigma}$ as  $\p{\sigma^\struct}{k}$.

% that equals \struct on static symbols and that interprets $\proj{\sigma}$ by 
% $\p{\sigma^\struct}{k}$ for dynamic symbols $\sigma$.

\end{definition}

%Given a tuple in \p{\Omega}{k}, one can reconstruct the
 % original tuple of $\Omega$. Hence: 

\begin{proposition}\label{prop:chains}
Let \voc be a linear-time vocabulary and $\Vss{\voc}$ the
  corresponding single state vocabulary.
  There is a  one-to-one correspondence between 
   \voc-structures \struct and sequences $(\struct_{k})_{k=0}^{\infty}$ of $\Vss{\voc}$-structures  sharing the same interpretation of static symbols, given by $J_k= \pss{J}{k}$.

\end{proposition}

This proposition holds because one can reconstruct the
  tuples in $\Omega$ from the  tuples in \p{\Omega}{k}. 
From now on, we will often identify a structure with the corresponding
 sequence. Often, we are not only interested in single states, but also in two successive states. 
 In the following definition, a \Vbs{\voc}-structure describes two
 subsequent states; $\proj{\sigma}$ refers to the first one and
 $\pron{\sigma}$ to the second.
% Therefore, we define the following concepts:

\begin{definition}[Bistate vocabulary and structure]
	For every dynamic symbol $\sigma$ in \voc, the \emph{next-state symbol} is a new symbol $\pron{\sigma}$ with the same type signature as \proj{\sigma}. 
The \emph{bistate vocabulary} \Vbs{\voc} extends
   the single-state vocabulary \Vss{\voc} with the symbol $\pron{\sigma}$ for each dynamic symbol. 
\ignore{
Given a \voc-structure $\struct$, the \emph{bistate projection}  $\pbs{\struct}{k}$ is the $\Vbs{\voc}$-structure equal to $\pss{\struct}{k}$ on $\Vss{\voc}$ and interpreting $\pron{\sigma}$ by $\p{\sigma^\struct}{k+1}$ for dynamic symbols $\sigma$.
}
With $\struct$ a \voc-structure, the \emph{bistate projection}
$\pbs{\struct}{k}$ over  $\Vbs{\voc}$ interprets $\pron{\sigma}$ as
$\p{\sigma^\struct}{k+1}$ and all other symbols as in $\pss{\struct}{k}$.
If $S$ and $S'$ are \Vss{\voc}-structures that are equal on static symbols, we use $(S,S')$ for the $\Vbs{\voc}$-structure with the same interpretation of static symbols and such that $\proj{\sigma}^{(S,S')}=\proj{\sigma}^S$ and $\pron{\sigma}^{(S,S')}=\proj{\sigma}^{S'}$ for dynamic symbols $\sigma$.
\end{definition}

%\newpage
\subsection{Progression and the Markov Property}
\begin{definition}[$k$-chain, extension]
Let $k$ be a natural number. A \emph{$k$-chain} $\struct$ over $\voc$ is a sequence $(\struct_{i})_{i=0}^{k}$ of $\Vss{\voc}$-structures sharing the same interpretation of static symbols.
Slightly abusing notation, we also call a \voc-structure an
$\infty$-chain (using the identification in  Proposition~\ref{prop:chains}).

The \emph{direct extension} of a $k$-chain \struct with a $\Vss{\voc}$ structure $S$, denoted $\extension{\struct}{S}$, is the $(k+1)$-chain $\struct'$ with $\struct'_j=\struct_j$ for $j\leq k$ and  $\struct'_{k+1}=S$. 
A chain $J''$ is an extension of $J$ if it is either a direct extension of $J$ or an extension of a direct extension of $J$.
\end{definition}
% % Abusing notation, we also call a \Vs{\voc}-structure a $-1$-chain and ; the above precision order is extended straightforwardly to these chains.
% \marginpar{deze paragraaf is niet essentieel}
% Using the identification of a structure with an $\omega$-chain, a $k$-chain is in fact a partial, or three-valued, structure, as defined in \cite{DeneckerBV12}.
% % It is a structure containing only information about an initial segment of the evolution of the state of affairs. %\todo{is dit stuk noodzakelijk?}
% %
% Under this identification, a chain $J''$ is an extension of $J$ if and only if the partial structure $J$ is less precise than $J''$.  

\begin{definition}[\theory-compatible, \theory-successor ]\label{def:tcompatible}
Let \theory be a \voc-theory.
A $k$-chain $\struct$ is \emph{\theory-compatible}  if \theory has a model (an
  $\omega$-chain) that extends \struct. 
A \Vss{\voc}-structure $S'$
  is a \emph{\theory-successor} of a $k$-chain \struct if \extension{\struct}{S} is \theory-compatible.
\end{definition}

\begin{definition}[Progression inference]\label{def:progression}
 \emph{Progression inference} is an inference that takes as input
 a theory \theory and a \theory-compatible $k$-chain \struct and
 returns all \theory-successors of \struct.
\end{definition}

% The structures $S$ returned by progression are always
% \theory-successors of the last \Vss{\voc}-structure of \struct.  
Of
special interest is the case where the \theory-successors of a $k$-chain \struct are determined solely by 
the last state in \struct. It means that the dynamic
system has no history. From a practical point of view this is often important. For example,
contemporary databases are often too large to keep track of the entire history. 
We refer to such a system as a system that has
the Markov property \cite{markov1906}.\footnote{The Markov property is
  often used in a probabilistic context. Translated to that context,
  one could say that the \theory-successors of $\struct$ are the states with
  non-zero probability.}  
\begin{definition}[Markov property]
	A theory \theory satisfies the \emph{Markov property} if for every  \theory-compatible $k$-chain $\struct$, and every  \theory-compatible $k'$-chain $\struct'$ ending in the same state, i.e., such that $\struct_k=\struct'_{k'}$, the \theory-successors of \struct are exactly the \theory-successors of $\struct'$.
\end{definition}

% 
% For a $k$-chain \struct to be \theory-compatible, it must
%   have an extension that is a model. As a consequence, computing a
%   progression $S$ of a chain \struct requires to look into the future
%   to check whether $\extension{\struct}{S}$ can be extended into a
%   model. We come back to this issue once we have introduced our Linear
%   Time Calculus.
%   
%   \ignore{
%TODO: dit is al een toekomstige tekst

The condition on a $k$-chain \struct to be \theory-compatible is quite strong. It does not only require that all information in this chain is correct according to \theory, but it requires that \struct is extensible to a model of \theory. This might require to look into the future. 
For example in the \pacman game, we could add two constraints: i) agents can not turn back and ii) as the game is not over, every agent moves. 
These two sentences are contradictory when an agent arrives at the end of a dead-end corridor. 
This means that every $k$-chain in which an agent \emph{enters} a
dead-end corridor is not \theory-compatible, as the agent will
eventually reach the point where it cannot move.
On the one hand, this is a good property, because progression as defined above guarantees that you can never get stuck, that every \theory-compatible chain can always be progressed.
But on the other hand, from a computational point of view, this is bad, as progression requires to look arbitrarily far into the future. 
In Appendix A, we present the notions of \emph{weak \theory-compatibility}, \emph{weak progression} and the \emph{weak Markov property} which are more technical than the one we described here. 
Intuitively, these properties are similar to the ones described here, except they do not require looking into the future. 
This might results in \emph{deadlocks}: chains without successors.
We implemented the weak progression; to the best of our knowledge, all systems that implement progression actually implement weak progression.

%\newpage
\subsection{The Linear Time Calculus}

%%%%%%%%%%%%%%%%%%%%%%%%%%%%%%%%%%%%%%%%%%%%%%%%
%%%%%%%%%%%%%%%%%%%%%%%%%%%%%%%%%%%%%%%%%%%%%%%%
%LTC-theoRIES
%%%%%%%%%%%%%%%%%%%%%%%%%%%%%%%%%%%%%%%%%%%%%%%%
%%%%%%%%%%%%%%%%%%%%%%%%%%%%%%%%%%%%%%%%%%%%%%%%

\begin{definition}[static, single-state, bistate]
\label{def:static}\label{def:initial}\label{def:bistate}\label{def:singlestate}
Let $\varphi$ be either a sentence or a rule.
 We call $\varphi$ \emph{static} if it contains no terms of type \time. We call $\varphi$ \emph{initial}  if it contains the constant \start and no other terms of type \time. 
 We call $\varphi$ \emph{single-state} if it contains a variable of type \time and no other terms of type \time. 
 We call $\varphi$ \emph{bistate} if it contains a variable $t$ typed \time and all terms typed \time in $\varphi$ are either $t$ or $\next(t)$. 
Furthermore, we call a single-state or a bistate $\varphi$ \emph{universal} if it is of the form $\forall t: \varphi'$ where $t$ is the unique \time-variable in $\varphi$ and $t$ is not quantified in $\varphi'$.
\end{definition}
 %quantified universally outside of the scope of negations and/or existential quantifications.
We now define an LTC theory as a theory that roughly only consists of the above types of rules and formulas. In this definition, we use the notion of stratification over \time: a definition is stratified over \time if it does not contain any rules defining atoms in terms of future values.
\begin{definition}[LTC-theory]\label{def:ltctheo}
An \emph{LTC-theory}
%\footnote{The ideas presented here are not inherently linked to \foid, LTC could also be defined for other logics. Therefore, we sometimes use \foid-LTC to emphasise the logic we start from.} 
over $\voc$ is a theory $\theory$ that satisfies i) all sentences and rules in \theory are either static, initial, universal single-state, or universal bistate, 
and ii) the definition in \theory is stratified over \time.%\footnote{This means that for all bistate rules in \theory, $\next(t)$ occurs in the unique dynamic position of the defined predicate in the head and that all single-state rules contain the time variable in the head.}
% \end{itemize}
\end{definition}
% \todo{RK: stratified over time is niet gedefinieerd}

The first condition ensures that an LTC-theory has no history.
For pure FO theories, this is enough to guarantee the Markov-property. 
The second condition prevents nonsensical definitions such as for example defining the state in terms of a future state. % at $\next(t)$.

\begin{example}
Below is an LTC theory over $\vocp$ (Example \ref{ex:pacmanvoc}) specifying part of the \pacman game\footnote{The complete example can be found at \citeN{url:PacManID}.}.
	\begin{ltheo}
	 \begin{ldef}
	   \forall a, p: \Position(a,\start) = p\lrule StartPos(a)=p.\\
	  \forall a, t, p: \Position(a,\next(t))=p\lrule \Position(a,t)=p\land \lnot \exists d: Move(a,d,t).\\
	   \forall a, t, p: \Position(a,\next(t))=p\lrule\exists d:  Move(a,d,t)\land Next(\Position(a,t),d,p).\\
	   \forall s: Pell(s,\start).\\
	   \forall s, t: Pell(s,\next(t))\lrule Pell(s,t) \land \Position(pacman,t)\neq s.\\
	 \end{ldef}\\
	\forall a, t, d, d': Move(a,d,t) \land Move(a,d',t)\limplies d=d'.
	 
	\end{ltheo}
The theory inductively defines the positions of the agent and the pellets at
  each time point (in terms of the open predicates $Next$, $StartPos$ and
  $Move$) and states the constraint that there is only one move at a time.

\end{example}

We now show how an LTC-theory can be translated automatically into two simpler theories: a \Vss{\voc}-theory that describes valid initial states, and a \Vbs{\voc}-theory that describes valid transitions. %where $S'$ is a \theory-successor of $S$.
% We will use the symbols $\proj{\sigma}$ for the first and $\pron{\sigma}$ for second of two successive states.

\begin{definition}[Elimination of time]
	Let $\varphi$ be a universal single-state or bistate sentence or rule with unique \time variable $t$. 
	The \emph{time-elimination} of $\varphi$ is the sentence/rule $\transl{\varphi}$ obtained from $\varphi$ by 
	(i) dropping the universal quantification of $t$,
	(ii) for every occurrence of $\next(t)$ in a dynamic (predicate or function) symbol $\sigma$, replacing $\sigma$ by $\pron{\sigma}$ and dropping the argument $\next(t)$, and 
	(iii) for every occurrence of $t$ in a dynamic  symbol, replacing $\sigma$ by $\proj{\sigma}$ and dropping  $t$.	
\end{definition}
% \begin{example}
	For example, the time-elimination of the rule 
\[\forall a, t, p: \Position(a,\next(t))=p\lrule\exists d:  Move(a,d,t)\land Next(\Position(a,t),d,p).\] is the \Vbs{\voc}-rule \[\forall a, p: \pron{\Position}(a)=p  \lrule\exists d:   \proj{Move}(a,d)\land Next(\proj{\Position}(a),d,p).\]
% \end{example}

\begin{defn}[Initial and  transition theory]\label{def:transfo}
	Let \theory be an LTC-theory. 
	We define two theories. The \emph{initial theory} \inittheo consists of: %\emph{transition theory} \transtheo as follows.
% 	\inittheo consists of:
		\begin{itemize}
			\item all static sentence/rules in \theory,
			\item for each initial sentence/rule $\varphi$ in \theory, the sentence/rule \transl{\forall t: \varphi[\subs{t}{\start}]} (informally, here we replace $\start$ by $t$ and project on $t$ afterwards since $te$ is only defined for universal sentences; this is the same as projecting on \start),
			\item for each single-state sentence/rule $\varphi$ in \theory, the sentence/rule \transl{\varphi}.
		\end{itemize}
	The \emph{transition theory} \transtheo consists of:
		\begin{itemize}
			\item all static sentences/rules in \theory,
% 			\item for each initial formula/rule $\varphi$ in \theory, the formula/rule \transl{\varphi[\subs{t}{\start}]}
			\item for each single-state sentence/rule $\varphi$ in \theory, the sentences/rules \transl{\varphi} and  \transl{\varphi[\subs{\next(t)}{t}]},  
			\item for each bistate sentence/rule $\varphi$ in \theory, the sentence/rule \transl{\varphi}.
		\end{itemize}
	
\end{defn}

\begin{example}
For our \pacman example, this results in the initial theory \inittheo:
	\begin{ltheo}
	 \begin{ldef}
	   \LRule{\forall a, p: \proj{\Position}(a) = p}{StartPos(a)=p}\\
	   \LRule{\forall s: \proj{Pell}(s)}\\
	 \end{ldef}\\
	\forall a, d, d': \proj{Move}(a,d) \land \proj{Move}(a,d')\limplies d=d'.
	\end{ltheo}
	and the transition theory \transtheo:
	\begin{ltheo}
	 \begin{ldef}
	   \forall a, p: \pron{\Position}(a)=p\lrule\proj{\Position}(a)=p\land \lnot \exists d: \proj{Move}(a,d).\\
	   \forall a, t, p: \pron{\Position}(a)=p\lrule\exists p', d: \proj{\Position}(a)=p' \land  \proj{Move}(a,d)\land Next(p',d,p).\\
% 	   \LRule{\forall a, p: \Position(a,\start) = p}{StartPos(a)=p}\\
% 	   \LRule{\forall s: Pell(s,\start)}\\
	   \forall s, t: \pron{Pell}(s)\lrule \proj{Pell}(s) \land \lnot \proj{\Position}(pacman)=s.\\
	 \end{ldef}\\
	\forall a, d, d': \proj{Move}(a,d) \land \proj{Move}(a,d')\limplies d=d'.\\
	\forall a, d, d': \pron{Move}(a,d) \land \pron{Move}(a,d')\limplies d=d'.
	 
	\end{ltheo}
\end{example}

% As is obvious, there is a strong correspondence between \theory and its two derived theories. 
% Informally, \inittheo and \transtheo contain enough information to characterise the models of \theory.
We now formalise the relation between \theory, \inittheo, and \transtheo; proofs can be found in Appendix~B.
% Informally, it is the case that models of \theory are exactly those structures \struct whose restriction to time $0$ is a model of \inittheo, and such that for every $n\in\Nat$, the restriction of \I to $n,n+1$ is a model of \transtheo. 
% 
% In order to formalise the above, we define some mappings between structures.

%%%%%%%%%%%%%%%%%%%%%%%%%%%%%%%%%%%%%%%%%%%%%%%%%%%%%
%%THEOREM
%%%%%%%%%%%%%%%%%%%%%%%%%%%%%%%%%%%%%%%%%%%%%%%%%%%%%%

\begin{theorem}\label{thm:main}
 Let \theory be an LTC-theory and $\struct$ a \voc-structure. 
 Then \struct is a model of \theory if and only if $\pss{\struct}{0} \models \inittheo$ and for every $k\in \Nat$, $\pbs{\struct}{k} \models \transtheo$.
\end{theorem}

\begin{theorem}\label{thm:weakmain}
 Let \theory be an LTC-theory and $\struct$ a $k$-chain. 
 Then, \struct is weakly \theory-compatible if and only if 
 $\pss{\struct}{0}\models  \inittheo$ and for every $j<k$, $\pbs{\struct}{j}\models \transtheo$.
\end{theorem}

\begin{corollary}\label{cor:markov}
 LTC-theories satisfy the Markov property and the weak Markov property.
\end{corollary}

%\newpage
\subsection{Modelling Methodology: Actions, Fluents and Inertia}\label{ssec:causes}
In many action languages, one often divides dynamic predicates into two sets of predicates: \emph{action predicates} and \emph{fluents}. 
Ever since the frame problem was defined by \citeN{McCHay69}, it has been clear that it is often easier to express state changes than it is to express the complete next state. 
Furthermore, from a practical standpoint, it is often easier to update a state---for example, a database---than it is to compute the entire next state. 
% Therefore, when modelling
Therefore, when modelling in LTC we often introduce  three extra predicate symbols for each fluent $P/n$: $C_P/n$, expresses that $P$ is caused to be true at a certain time, $C_{\lnot P}/n$ expresses that $P$ is caused to be false and $I_P/(n-1)$ expresses that $P$ holds initially. 
The relation between these predicates is formalised in LTC:
\begin{ltheo}
 \begin{ldef}
\LRule{\forall \xxx: P(\xxx,\start)}{I_P(\xxx)}\\
   \LRule{\forall \xxx, t: P(\xxx,\next(t))}{C_{P}(\xxx,t)}\\
  \LRule{\forall \xxx, t: P(\xxx,\next(t))}{P(\xxx,t)\land \lnot C_{\lnot P}(\xxx,t)}
 \end{ldef}
\end{ltheo}
I.e., $ P(\xxx,\next(t))$ holds if it is either caused to be true or it was already true and is not caused to be false (inertia). 
Using this methodology, the modeller simply describes the effects of actions through $C_P$ and $C_{\lnot P}$ and a reasoning engine can exploit these new predicates for efficiently updating a persistent state.

% \section{Progression}
%   \input{03-progression}

\section{LTC-Theories in Practice: Inferences and Implementation}\label{sec:inferences}
  This section describes various inference methods we can use on LTC-theories with a brief description of their implementation in \idpthree, available in version $3.3$ \cite{url:idp}.

\subsection{Progression}
Based on Theorem~\ref{thm:weakmain}, we can use model expansion
  on the initial theory to infer an initial state 
%   (the interpretation  of the $\proj{\sigma}$ symbols) 
and model expansion
  on the transition theory to infer a next state 
%   (the interpretation  of the $\pron{\sigma}$ symbols) 
from a given
  state. 
%   (the interpretation of the $\proj{\sigma}$ symbols is in the   input structure). 
 To perform these
  inferences, we added  two procedures to \idpthree:

\begin{itemize} 
% \item
 \item \texttt{initialise(T,J)} takes as input an LTC-theory $T$ over \voc and a partial \voc-structure $J$ (a structure that at least interprets all types) and returns a set of \Vss{\voc}-structures that are initial states of $T$ and that agree with $J$ (that expand $\pss{J}{0}$).
 The number of generated \Vss{\voc}-structures depends on the option \texttt{nbmodels}.
 \item \texttt{progress(T,S)} takes as input an LTC-theory $T$ over \voc and any  \Vss{\voc}-structure $S$ and returns a set of \Vss{\voc}-structures $S'$ for which $(S,S')\models T_t$. 
The number of generated \Vss{\voc}-structures depends on the option \texttt{nbmodels}.
\end{itemize}
% \old{plemented these inference by first applying the transformation described in Definition \ref{def:transfo} and subsequently using the existing model expansion inference.}
% We cache the transformed theories for efficiency purposes.
 
\subsection{Logic-Based Software Development using Interactive Simulation}
An LTC-theory describes the evolution of a dynamic system over time.
By itself, it cannot interact with the external world.
In order to create such software, we can interactively simulate an LTC-theory by waiting for user input at each progression step.
The simplest form is a procedure that uses the progression inference to present all possible next states to a user, and asks to pick one. 
% This is the form of interactive simulation available in the NuSMV system \cite{CAV/CimattiCGGPRST02}. 
We implemented this form of interactive simulation in \idpthree:
calling \texttt{simulate\_interactive(T,J)}, with a \voc-LTC-theory $T$ and a partial \voc-structure $J$ at least interpreting all types in \voc, provides you with an interactive shell to guide the simulation.  
Interactive simulation reuses the \texttt{initialise} and \texttt{progress} procedures.
Due to its rather primitive communication with the user, this kind of simulation is not yet useful for running software based on a logical specification. 
In order to do so, we need a more refined form of interaction. 

\subsubsection{Modelling Methodology: Exogenous and Endogenous Information}
One way to achieve a more refined form is by making an explicit distinction between \emph{exogenous} and \emph{endogenous} information, respectively information determined by the environment and information internal to the system.
In most applications, fluents are endogenous and (a subset of the) actions are exogenous.
For \pacman, the only exogenous information is the action the player takes: the direction in which he moves.
In order to  allow such a distinction (and in the meantime, many other refined control mechanisms), we implemented \texttt{simulate($T$, $J$, rand, show(), endcheck(), choose())}. 
Besides an LTC-theory $T$ and a partial structure $J$, this inference takes as input:
\begin{itemize}
%  \item $T$: a \voc-LTC-theory,
%  \item $J$: a \voc-structure at least interpreting all types in \voc,
 \item {\tt rand}: a boolean; if true, simulations happens randomly, otherwise interactively,
 \item (optional) {\tt show()} a Lua-procedure that implements printing of the current state,
 \item (optional) {\tt endcheck()} a Lua-procedure that decides whether to stop the simulation, 
 \item (optional) {\tt choose()} a Lua-procedure that implements choosing a next state.
\end{itemize}
This procedure also simulates $T$, but all communication goes through the user provided procedures. 
It reuses the \texttt{initialise} and \texttt{progress} procedures.
We used the above procedure to simulate a game of \pacman.
As {\tt show} procedure, we passed a call to the visualisation tool \idpdraw~\cite{url:idpdraw}; the stop-criterion checks for the atom ``$GameOver$'', and our {\tt choose} procedure asks the user which direction to go to. 
This results in a complete, playable \pacman implementation that can be found at~\citeN{url:PacManID}. At the moment, behaviour of the ghosts is random, but this could easily be replaced by smart AI by providing a specification for the behaviour of the ghosts.

% \todo{iets over BLoxUI?}

\subsection{Proving Invariants}\label{ssec:invariants}
\begin{definition}
  An \emph{invariant} of an LTC-theory \theory is a universal single-state sentence $\varphi$ such that $\theory\models\varphi$. 
\end{definition}
The straightforward way to prove invariants is theorem proving (deduction inference). % and hoping to get a satisfactory answer (deduction inference).\mbart{deduction bijgekomen, neemt plaats}
In \idpthree, this can be done using the procedure \texttt{entails(T,f)}, which checks whether sentence  $f$ is entailed by theory  $T$. 
\idpthree automatically translates this call to a theorem prover supporting TFA \cite{conf/lpar/SutcliffeSCB12} or FOF \cite{Sutcliffe09}. 
% The option \texttt{provercommand} controls which theorem prover is used.
Often, theorem provers are unable to prove entailed invariants. This can happen for example because the nature of \time (\nat) is not exploited enough or because this problem is undecidable in general.
The following theorem shows that for LTC-theories, we can prove invariants by induction.
\begin{theorem}\label{thm:entailment}
 Let \theory be an LTC-theory and $\varphi$ a universal single-state sentence. 
 Then $\theory\models\varphi$ if $\inittheo \models \transl{\varphi}$, and $(\transtheo \land \transl{\varphi} ) \models \transl{\varphi[\next(t)/t]}$, where $t$ is the unique time-variable in $\varphi$.
\end{theorem}
%  This theorem is in fact a reformulation of the principle of proofs by induction.
 We added a procedure \texttt{isinvariant(T,f)} to the \idpthree system; this procedure checks the two entailment relations from Theorem \ref{thm:entailment}. 
 It uses the same transformations as the progression inference and reuses the existing deduction  inference of \idpthree.

\subsubsection{Fixed-Domain Invariants}
Some sentences are not invariants in general, but only in a certain context (i.e., in a given domain). 
For example: in a given \pacman grid it might be that \pacman can always reach all remaining pellets. 
This is however not a general invariant of the \pacman game since it is possible to construct grids in which some pellets are completely surrounded by walls.
In case a finite domain for all other types than \time and an interpretation for some static symbols is given, it suffices to search for counterexamples of the invariant in this specific setting.
% However, in general, not all squares might be connected.
\begin{theorem}\label{thm:invariantMX}
	 Let $\struct$ be a $\Vs{\voc}$-structure and let $\varphi$ be a universal single-state sentence with time variable $t$.
	 Then $\varphi$ is satisfied in all \voc-structures expanding $\struct$ if $\inittheo \land \lnot \transl{\varphi}$ has no models expanding $\struct$, and 
		 $\transtheo \land \transl{\varphi}  \land \lnot  \transl{\varphi[\next(t)/t]}$  has no models expanding $\struct$.
\end{theorem}
% Again, this lemma follows immediately using the principle of proofs by induction. 
We added the procedure \texttt{isinvariant(T,f,J)} to \idpthree; this procedure checks whether sentence $f$ is an invariant of  $T$ in the context of $\struct$ using Theorem \ref{thm:invariantMX}. 
 It reuses the transformations implemented for progression and the model expansion inference of \idpthree.

 \subsubsection{More General Properties}
 
 Proving invariants is often useful. But in many cases one is interested in proving more general properties. 
 For example in the \pacman game, a desired property would be that pellets never reappear: $\forall t, s: \lnot  Pell(s,t) \limplies \lnot Pell(s,\next(t))$.
 This sentence is a universal bistate sentence. For these formulas, we find a result similar to Theorem \ref{thm:entailment}. 
 \begin{theorem}\label{thm:bistateinvar}
 Let \theory be an LTC-theory and $\varphi$ a universal bistate sentence. 
 If $\transtheo \models \transl{\varphi}$ then $\theory\models\varphi$.
 \end{theorem}
The above theorem not only yields a method to prove bistate invariants, but also a method to prove them in the context of a given domain similar to Theorem \ref{thm:invariantMX}. 
The procedures  \texttt{isinvariant(T,f)} and  \texttt{isinvariant(T,f,J)} automatically detect whether sentence  $f$ is a bistate or a single-state invariant and apply the appropriate methods.
For proving more complex properties $\varphi$, the only method available yet is directly proving that  $\theory\models\varphi$. %Sometimes, one even wants to verify properties 
% \mmaurice{wat vb. van pacman properties die (niet) bewezen konden worden?}

\subsection{Planning}
\ignore{In the context of a dynamic domain, one important computational task is \emph{planning}: finding a sequence of actions reaching a certain goal state.
In \idpthree, in order to do this, one typically creates a second theory describing the goal state.
For example,  a goal state could be the condition that \pacman wins, 
$	\exists t: \forall s: \lnot Pell(s,t).$
Searching a plan is then a simple application of model expansion inference  (after merging the LTC-theory with the goal theory).
In the standard setting, this requires all domains (including \time) to be finite, but recent work on Lazy Grounding sometimes removes this restriction \cite{iclp/DeCatDS12}.
}

For dynamic domains, \emph{planning} is an important computational
task: finding a sequence of actions reaching a certain goal state. 
To do this in \idpthree, one typically creates a second theory
describing the goal state. 
As an example, the condition that \pacman wins, $\exists t: \forall s: \lnot Pell(s,t)$, is a goal state. 
A plan can then be searched through model expansion inference, after merging the LTC-theory with the goal theory.
In the standard setting, this requires all domains (including \time) to be finite, but recent work on Lazy Grounding removes this restriction \cite{iclp/DeCatDS12}.

Often, a cost is associated with each plan, e.g., the number of steps needed to win the game.
The minimisation inference in \idpthree searches for a plan with minimal cost.

\section{Related Work}\label{sec:comparison}
% \todo{a comparison with the interactive system
%   oclingo would be nice,
%   although I'm not sure how this can be compared.}
% 
% \todo{ I would have liked to see stronger justifications for the missing ticks in Table 1. For example, [1] shows how ASP can in fact be used to prove invariants based on an action language-like specification of dynamic systems. [1] S. Haufe, S. Schiffel, M. Thielscher. Automated verification of state sequence invariants in general game playing. AIJ 187-188:1-30, 2012.}

Many action languages are closely related to LTC; the relation between several of these calculi has been studied intensively by \citeN{ai/Thielscher11}.
The focus of this paper is not on the language, but on the forms of
inference we can perform. % on knowledge expressed in our language. 
In what follows, we discuss several domains and systems concerned with inference for (dynamic) languages.
We show that \idpthree distinguishes itself by the variety of inferences it offers.  
An overview of the discussed domains and systems can be found in Table \ref{tab:comparison}. 

\newcommand{\YES}{\m{\times}}
\newcommand{\NO}{\m{-}}
\begin{table}[htbf]
\begin{center}
\begin{tabular}{@{}l|@{}c@{}c@{}c@{}c@{}c@{}c@{}c@{}c@{}}
			& \idpthree 	& DS 	& Pl 	& TP 	& ASP-CP	& NuSMV	& LPS 	& ProB				\\
 \cline{1-9}
Progression 		& \YES 		& \YES 	& \NO 	& \NO 	& \NO		& \YES	& \YES	& \YES			\\
Planning 		& \YES 		& \NO	& \YES	& \NO 	& \YES		& \YES	& \YES	& \YES 			\\
Optimal Planning 	& \YES 		& \NO	& \YES	& \NO 	& \YES		& \NO	& \NO	& \YES 			\\
Proving Invariants	& \YES		& \NO	& \NO	& \YES	& \NO 		& \NO	& \NO	& \NO 			\\ 
Interactive Simulation	& \YES		& \YES	& \NO	& \NO	& \NO		& \YES	& \YES	& \YES 			\\

LTL/CTL Model Checking	& \NO	& \NO	& \NO	& \NO	& \NO		& \YES	& \NO	& \YES

\end{tabular} 
\end{center}
\caption{The various inferences (rows) and systems/fields (columns) we consider in this comparison:  \idpthree, Database Systems (DS), Planners (Pl), Theorem Provers (TP), ASP/CP-solvers (ASP-CP), NuSMV, LPS and ProB.
  }
\label{tab:comparison}
\end{table}

% 
% Many other systems implement similar inferences to the ones described here. 
% The
% Many of the systems and research domains discussed below only implement or study one form of inference. 
% Hence, many of them will be more efficient at that specific task. 
% But, as this paper shows, in many contexts, being able to reuse your specifications, your knowledge, for multiple inference tasks yields a great advantage. 

Many database systems implement some form of progression  \cite{journals/ai/LinR97}.
Often, these systems use (a variant of) transaction logic \cite{BonnerKC93} to express  progression steps. %, or transactions, as they are often called.
% For example, transactions in the LogicBlox system \cite{datalog/GreenAK12} contain rules of the form $p\lrule \varphi$, where $\varphi$ is a formula. In this case, $p$ refers to the next state, while $\varphi$ can refer either to the current or the next state, and this is not specified. \todo{uitleggen hoe dat dan wel werkt?}
% Mose of those database systems, only support one inference on a ``dynamic'' theory, namely progression. 
Other dynamic inferences, such as backwards reasoning, planning, and
verification are, to the best of our knowledge, not possible in these
systems.  %\todo{query inferentie ondersteunen ze ook}
A very interesting  database system is LogicBlox
\cite{datalog/GreenAK12}; it supports a refined interactive simulation
by means of a huge set of built-in predicates (windows, buttons,
etc.). Users can specify workflows  declaratively; during simulations,
the UI is derived from  the interpretations of the built-ins.

Proving invariants can be handled by 
theorem provers such as SPASS \cite{cade/WeidenbachDFKSW09}, Vampire \cite{aicom/RiazanovV02}, and many more.
\citeN{Sutcliffe13} desbribed an overview of state-of-the-art theorem provers. 
Provers are only able to handle one form of inference, namely deduction.
% In fact, \idpthree simply translates entailment calls to any theorem prover 
% supporting TFA \cite{conf/lpar/SutcliffeSCB12} or FOF \cite{Sutcliffe09}. 
We optimised this for the case of proving invariants of an LTC-theory using  induction. 
Some interactive theorem provers, for example ACL2 \cite{KaufmannSP00} and Coq \cite{misc/Coq:manual}, can generate inductive proofs but they require guidance from the user.

Another community with great interest in dynamic specifications is the \emph{planning} community.
Many planners support the PDDL language \cite{ghallabGCPSSSSD98};
\citeN{AmandaAGOJLSY12} published an overview of such systems. 
To the best of our knowledge, these systems only support one form of inference, namely planning. 
The planning inference is in fact a special case of model expansion. 
This is demonstrated for example by a tool that translates PDDL specifications into LTC-theories \cite{url:pddl2idp}.
Planning problems can also be encoded in other systems that
essentially perform model expansion, such as Answer Set Programming (ASP)
\cite{GelfondL98} systems, or using
constraint programming \cite{Apt03} or
 (integer) linear programming \cite{NW:icopt}.
In ASP, systems that perform other inferences on dynamic systems have been developed as well. For example oClingo \cite{kr/GebserGKOSS12} allows stream reasoning, a form of interactive simulation and \citeN{journals/ai/HaufeST12} describe methods to prove invariants of (temporal) ASP encodings of game specifications. However, these various methods have not been unified in one system to work on the same specification.
% For each of these domains many solvers exist, including incremental ASP systems, such as iClingo \cite{iclp/GebserKKOST08}. 
% These domains essentially all solve the same problem, model expansion, even though using different terminologies. 

The above discussion focuses on general fields tackling (only) one of the problems we are typically interested in (in a dynamic context).
To be fair, it is worth mentioning that systems in these domains often tackle a more general problem (for example ASP systems can do much more than only planning). %, and theorem provers can also prove general statements, not only invariants). 
 \idpthree tackles these more general problems efficiently as well. 
Over the years, \idp %(\idpthree and its predecessors) 
and \minisatid (the solver underlying \idpthree) have proven to be among the best ASP and CP systems \cite{LPNRM/Calimeri11,conf/lpnmr/AlvianoCCDDIKKOPPRRSSSWX13,misc/amadini2013}.

Many other systems are designed for dynamic domains. 
For example, NuSMV \cite{CAV/CimattiCGGPRST02} supports progression, interactive simulation, CTL and LTL model checking, and planning (by giving counterexamples for the LTL statement that the goal cannot be reached). 
This system is propositional, hence symbolic, domain-independent
proving of invariants is impossible.
CTL and LTL model checking are currently not supported by
  \idpthree. 
 However, conceptually they form no problem: LTL properties can be translated into \voc-sentences and deduction inference could be used to prove them. 
 Furthermore, progression inference could be used to generate a state graph, on which more efficient CTL and LTL and model checking algorithms can be applied. This is not yet implemented.

The LPS framework from \citeN{DBLP:journals/corr/KowalskiS13}
  has a lot of goals in common with our work, it aims at providing a unified
  framework for computing with dynamic systems. The language is richer
  than LTC as rules can relate more than two points in time and there
  is an explicit representation of external events. The model-theoretic
  semantics is pretty  close to the \foid semantics. The operational
  semantics corresponds to simulation: it works on time-eliminated states and selects a single successor
  state. Similar to weak progression, it cannot look in the future, and hence might result in a deadlock. The current
  implementation is on top of Prolog and mainly aims at (interactive)
  simulation.

% \todo{RK: execution, niet simulation}
% \old{In \cite{DBLP:journals/corr/KowalskiS13}, the LPS framework is presented as ``a logic-based unifying framework for computing''. 
% The language of LPS describes changes to a state, similar to the modelling methodology presented in Section \ref{ssec:causes}.
% In \cite{DBLP:journals/corr/KowalskiS13}, we read that ``computation in the framework is viewed as the task of generating a sequence of state transitions, with the purpose of making an agent's goals all true''. 
% This suggests a focus on planning above other computational tasks; planning is one of the inferences LPS can perform.
% The operational semantics of LPS programs corresponds to performing progression steps. 
% By interaction with other agents in the system, LPS also supports interactive simulation. 
% Inferences like model checking and proving invariants are not (yet) possible in this framework.}

The ProB system \cite{journals/sttt/LeuschelB08} is an automated animator and model checker for the B-Method. 
It can provide interactive animations (interactive simulation) and can also be used to do (optimal) planning and automatically verify dynamic specifications. 
ProB is a very general and powerful system. The only inference studied in this paper it does not support is domain independent proving of invariants.

% 
% 
% \begin{itemize}
% % 	\item database systemen. -> progressie query
% 	\item NuSMV en andere symbolic model checkers? -> de meeste enkel TODO REFERENTEIS UIT NUSMV
% 	\item Event-B 
% % 	\item Theorem provers ->  paradox
% % 	\item CP systemen, ASP systemen, ...  -> model generation, incremental model generation iclingo
% % 	\item Kowalski 
% 	\item SMT dingen voor model checking en verificatie, -> verschillende taal voor programma en verificatie anderzijds. 
% 	\item Flora, transactional logic, Kiefer
% % 	\item Planningsystemen -> 
% 
% \end{itemize}

\section{Conclusion and Future Work} \label{sec:conclusion}
In this paper we studied how the KBS paradigm can be applied in the context of dynamic domains. 
We identified many interesting forms of dynamic inference 
and explained how all of these inferences can be applied in the context of software development based on logic. 
We showed that in principle, each of these inferences can be applied on the same problem specification, and thus argued the importance of knowledge reuse. 

Furthermore, we implemented, with relative ease, all but one of these inference methods in \idpthree. 
The general approach consisted of translating inference tasks in the context of a dynamic domain to existing inference methods.
% This demonstrate the  the extensibility of \idpthree. 
Afterwards, we compared \idpthree to other formalisms and systems and conclude that 
\idpthree is one of the few systems supporting this much inferences on dynamic specifications.

Integrating CTL and LTL verification algorithms is %an interesting 
a topic for future work.

%\bibliographystyle{acmtrans}
%%
%\bibliography{krrlib}

 \appendix
 \newpage
 \section{Weak Progression}\label{app:weakprogression}
 In this section, we show how a weaker variant of progression can be defined using three-valued logic. 
We will restrict our attention to function-free vocabularies (i.e.,vocabularies containing only constant and predicate symbols) here to simplify the presentation. 
However, all definitions can be extended to the general case.

\subsection{Three-valued logic}

\newcommand\pset{\m{\mathcal{P}}}
\newcommand\assignvar[2]{#2:#1}

We briefly summarise some concepts from three-valued logic.
A truth-value is one of the following: $\{\ltrue,\lfalse,\lunkn\}$ (true, false and unknown). 
% The two-valued truth values are the standard Booleans $\bool=\{\ltrue,\lfalse\}$.
We define $\lfalse^{-1}=\ltrue, \ltrue^{-1}=\lfalse$ and $\lunkn^{-1}=\lunkn$.
We define two orders on truth values: the precision order 
 \leqp is given by $\lunkn\leqp\ltrue$ and $\lunkn\leqp\lfalse$.
 And the truth order $\leq$ is given by $\lfalse\leq\lunkn\leq\ltrue$.
\begin{definition}
	A \emph{partial set} \pset over $D$ is a function from $D$ to $\{\ltrue,\lfalse,\lunkn\}$.
\end{definition}

The precision order is extended to partial sets over $D$: $\pset\leqp \pset'$ if for all $d\in D: \pset(d)\leqp\pset'(d)$.

A partial \voc-structure \struct consists of 
1) a \emph{domain}, \domainof{\struct}: a set of elements, and
2)  a mapping associating a value to each symbol in \voc. 
For predicate symbols $P$ of arity $n$, this is a partial set $\inter{P}{\struct}$ over $(\domainof{\struct})^n$. For constants, this is a value in \domainof{\struct}.

We assume that a (partial) structure also interprets variable symbols and denote $\struct[\assignvar{d}{x}]$ for the structure equal to $\struct$ except interpreting $x$ by $d$.

If $\struct$ and $\struct'$ are two partial structures with the same interpretation for constants, $\struct$ is less precise than $\struct'$ ($\struct\leqp\struct'$) if for all symbols $\sigma$,
$\inter{\sigma}{\struct}\leqp\inter{\sigma}{\struct'}$.
A partial structure $\struct$ is two-valued if interpretations of its symbols map nothing to $\lunkn$. 
A two-valued partial structure is exactly a structure.

% From now on, we restrict our attention to partial structures that are a partial function from $\nat$ to \Vss{\voc}-structures, i.e.,

\begin{definition}\label{def:kleeneval}
Given a partial structure \struct, 
the Kleene valuation ($\kleeneval_{\struct}$) is defined inductively based  on the Kleene truth tables \cite{Kleene38}:
\begin{itemize}
	\item $\kleeneval_{\struct}(P(\ttt)) = P^{\struct}(\ttt^\struct)$,
	\item $\kleeneval_{\struct}(\neg\varphi) = (\kleeneval_{\struct}(\varphi))^{-1}$
	\item $\kleeneval_{\struct}(\varphi\land\psi) = \min_\leq\left(\kleeneval_{\struct}(\varphi),\kleeneval_{\struct}(\psi)\right)$
	\item $\kleeneval_{\struct}(\varphi\lor\psi) = \max_\leq\left(\kleeneval_{\struct}(\varphi),\kleeneval_{\struct}(\psi)\right)$
	\item $\kleeneval_{\struct}(\forall x:\varphi) = \min_\leq\left\{\kleeneval_{\struct[\assignvar{d}{x}]}(\varphi)\mid d\in\domainof{\struct}\right\}$ %\mid d\in \ptinter{\domain}{\struct}
	\item $\kleeneval_{\struct}(\exists x:\varphi) = \max_\leq\left\{\kleeneval_{\struct[\assignvar{d}{x}]}(\varphi)\mid d\in\domainof{\struct}\right\}$
\end{itemize}
\end{definition}

The Kleene valuation is extended to definitions and theories.
For definitions, intuitively, the value of \D is true if all its defined atoms are two-valued and have the correct (defined) interpretation, its value is false if some defined atom is interpreted incorrectly, and is unknown otherwise. 
The exact definition can be found in \cite{tocl/DeneckerT08}. In this text, we will only use the following property.
\begin{proposition}\label{prop:defunkn}
 If all defined atoms in a non-empty definition \D are interpreted as $\lunkn$ in $J$, then $\kleeneval_J(\D)=\lunkn$. 
\end{proposition}
We use $\kleeneval_\struct(\theory)$ to denote the Kleene value of a theory $\theory$ over a structure $\struct$. 
$\kleeneval_\struct(\theory) = \ltrue$ if all of $\theory$'s definitions and sentences have value \ltrue in structure $\struct$.
$\kleeneval_\struct(\theory) = \lfalse$  if one of $\theory$'s definitions or sentences has value \lfalse in structure $\struct$.
$\kleeneval_\struct(\theory) = \lunkn$ otherwise.

We summarise some well-known properties about the Kleene-valuation.
\begin{proposition}\label{prop:kleene2val}
 If $\struct$ is a two-valued partial structure (i.e., a structure), then $\kleeneval_\struct(\theory)$ is $\ltrue$ if and only if $\struct\models\theory$ and  $\kleeneval_\struct(\theory)$ is  $\lfalse$ otherwise.
\end{proposition}

\begin{proposition}\label{prop:kleeneprecise}
 If $\struct$ and $\struct'$ are partial structures with $\struct\leqp\struct'$, then for every theory \theory,
 $\kleeneval_\struct(\theory)\leqp\kleeneval_{\struct'}(\theory)$.
\end{proposition}

\subsection{Weakly \theory-Compatible Chains and Weak Progression}
For this paper, we are only interested in a special kind of partial structures: partial structures that have complete information on  an initial segment of time points and that have  no information about other time points. 
Using the identification of a structure with an $\infty$-chain, a $k$-chain corresponds to such a partial structure. 
If $(J_j)_{j=0}^{k}$ is a $k$-chain, we associate to $J$ the partial structure \struct equal to the $J_j$ on static symbols and such that for dynamic symbols $\sigma$
\[
 \sigma(d_1 \dots d_{n-1},j)^\struct = \left\{\begin{array}{l}
                                                             \ltrue \text{ if } j\leq k \text{ and }(d_1 \dots d_{n-1}) \in \proj{\sigma}^{J_j}\\
                                                             \lfalse \text{ if } j\leq k \text{ and }(d_1 \dots d_{n-1}) \not\in \proj{\sigma}^{J_j}\\
                                                             \lunkn \text{ otherwise}
                                                           \end{array}\right.
\]
We identify the $k$-chain and the corresponding partial structure. 

\begin{definition}[Weakly \theory-compatible, weak \theory-successor] 
A $k$-chain $\struct$ is \emph{weakly \theory-compatible} with a
  \voc-theory \theory if $\kleeneval_\struct(\theory)\neq\lfalse$.
  
  A \Vss{\voc}-structure $S'$
  is a \emph{weak \theory-successor} of a $k$-chain \struct if \extension{\struct}{S} is weakly \theory-compatible.
\end{definition}

\begin{proposition}\label{prop:compatibleweakcompatible}
 Every \theory-compatible $k$-chain $\struct$ is also weakly $\theory$-compatible.
\end{proposition}
 \begin{proof}
  If \struct is \theory-compatible, then there is a model $\struct'$ of \theory that is more precise than \struct. 
  Since $\struct'\models\theory$, $\kleeneval_{\struct'}(\theory)=\ltrue$ by Proposition \ref{prop:kleene2val}.
  Now, Proposition \ref{prop:kleeneprecise} guarantees that $\kleeneval_{\struct}(\theory)$ is less precise than $\ltrue$, hence it must be either $\ltrue$ or $\lunkn$ and we conclude that $\struct$ is indeed weakly \theory-compatible.
 \end{proof}
 
 The reverse of Proposition \ref{prop:compatibleweakcompatible}
 does not hold as the following (simple) example shows.
 
 \begin{example}
  Let \theory be the following first-order theory:
  \begin{ltheo}
   P(\start).\\
   \forall t:Q(\next(t))\lequiv P(t).\\
   \forall t: \lnot Q(t).
  \end{ltheo}
  It is clear that \theory has no models, as the second constraint requires $Q$ to be true at time $1$, while the last constraint requires $Q$ to be false at all time points.
  Hence, there are no \theory-compatible chains.
  
  However, the $0$-chain \struct such that $\struct_0$ interprets $P$ by $\ltrue$ and $Q$ by $\lfalse$ is weakly \theory-compatible. The Kleene-valuation of \theory in \struct is $\lunkn$.

 \end{example}

The above example shows that it is possible that a weakly \theory-compatible chain cannot be extended. Such a situation is often called a deadlock. 
\begin{definition}[Deadlock]
 A weakly \theory-compatible chain $\struct$ is in a \emph{deadlock} if there are no weakly \theory-compatible extensions of \struct.
\end{definition}

\begin{definition}[Weak Progression inference]\label{def:weakprogression}
 The \emph{weak progression inference} is an inference that takes as input
 a theory \theory and a weakly \theory-compatible $k$-chain \struct and
 returns all weak \theory-successors of \struct.
\end{definition}

\begin{definition}[Weak Markov property]
	A theory \theory satisfies the \emph{weak Markov property} if for every weakly \theory-compatible $k$-chain $\struct$, and every  weakly \theory-compatible $k'$-chain $\struct'$ ending in the same state, i.e., such that $\struct_k=\struct'_{k'}$, the weak \theory-successors of \struct are exactly the weak \theory-successors of $\struct'$.
\end{definition}

The weak Markov property essentially says the same as the Markov property, namely that the successors of a given chain only depend on the last state, i.e., that the system has no history.

 \newpage
 \section{Proofs}\label{app:proofs}
 %\todo{appendix voor of na refs?}

%\maurice{erna, de refs komen in de paper, de appendix alleen in
%  on-line supplement.} 

% \maurice{herhaal de formuleringen}

\noindent
\emph{Proposition~3.6}

\noindent
Let \voc be a linear-time vocabulary and $\Vss{\voc}$ the
  corresponding single state vocabulary.
Then the mappings $\pss{\cdot}{k}$ induce a one-to-one correspondence between 
   \voc-structures \struct and sequences $(\struct_{k})_{k=0}^\infty$ of $\Vss{\voc}$-structures  sharing the same interpretation of static symbols.

\begin{proof}%[Proof of Proposition \ref{prop:chains}]
 It is clear that given a structure $J$, $(\pss{J}{k})_{k=0}^{\infty}$ is indeed such a sequence. 
 
 Now, for the other direction, suppose $J_k$ is a sequence of $\Vss{\voc}$-structures sharing the same interpretation of static symbols.
  Let $J$ denote the \voc-structure with the same interpretation of static symbols and such that, for dynamic predicates $\sigma$, 
  \[\sigma^J= \{(d_1,\dots,d_{n-1}, k )\mid (d_1,\dots,d_{n-1}) \in \proj{\sigma}^{J_k}\}.\]
  Then $J$ is indeed a structure such that $\pss{J}{k}=J_k$, as desired. 
\end{proof}

\noindent
\emph{Theorem 3.18}

\noindent
 Let \theory be an LTC-theory and $\struct$ a \voc-structure. 
 Then \struct is a model of \theory if and only if $\pss{\struct}{0} \models \inittheo$ and for every $k\in \Nat$, $\pbs{\struct}{k} \models \transtheo$.

\begin{proof}%[Proof of Theorem 3.18]
By the first condition of Definition 3.13, the
  FO part of the theory
 only consists of static, initial, single-state, and bistate sentences.
 Now, a structure $J$ satisfies a static sentence if and only if each of its projections satisfy this sentence. A structure $J$ satisfies an initial sentence, if and only if its initial time-point satisfies the projection of this sentences, etc. Hence, for the FO part, the result easily follows. 
	
	Furthermore, Definition 3.13 guarantees that all definitions in \theory are stratified over time. 
	Now, it follows immediately from Theorem 4.5 in \cite{tocl/VennekensGD06} that we can split stratified definitions in one definition for each stratification level. 
	Thus, what we obtain is one definition for each point in time, defining the state at $\next(t)$ in terms of the state in $t$. 
	This definition corresponds exactly to the definition in \transtheo, as desired. 
\end{proof}

\noindent
\emph{Theorem~3.19}

\noindent
Let \theory be an LTC-theory and $\struct$ a $k$-chain. 
 Then, \struct is weakly \theory-compatible if and only if 
 $\pss{\struct}{0}\models  \inittheo$ and for every $j<k$, $\pbs{\struct}{j}\models \transtheo$.

\begin{proof}%[Proof of Theorem 3.19]
 One direction is clear: if $\struct$ is weakly \theory-compatible, then $\pss{\struct}{0}\models  \inittheo$ and for every $j<k$, $\pbs{\struct}{j}\models \transtheo$. 
 
 For the other direction, suppose  $\pss{\struct}{0}\models  \inittheo$ and for every $j<k$, $\pbs{\struct}{j}\models \transtheo$. We will show that $J$ is weakly \theory-compatible. In order to show this, we will show that $\kleeneval_J(\theory)\neq \lfalse$, or said differently, that for every sentence $\varphi \in \theory$, $\kleeneval_J(\varphi)\neq \lfalse$ and that for the definition \D in \theory, $\kleeneval_J(\D)\neq \lfalse$.
 
 First, let $\varphi$ be any sentence in \theory. 
 If $\varphi$ is an initial, or a static sentence, then $\struct\models\varphi$ because  $\pss{\struct}{0}\models  \inittheo$, thus $\kleeneval_J(\varphi)=\ltrue$ for such sentences.  
 If $\varphi$ is a universal single-state sentence $\forall t: \varphi'(t)$, we assume that $\kleeneval_\struct(\varphi)=\lfalse$, and will show that this leads to a contradiction. 
%
% \maurice{? je bedoelt dat je aanneemt dat  $\varphi$ false is in
%   \theory? ( $\theory \models \neg \varphi$?? (J is een k-chain) in het ander geval heeft J een extensie
%   die een model is en volgt de claim uit de vorige stelling??}
%
In this case, using the definition of the Kleene valuation, at least for one $i$,
$\kleeneval_\struct(\varphi[\subs{i}{t}])=\lfalse$, or said
differently, at least for one $i$,
$J_i\not\models\transl{\varphi}$. Now, this $i$ should definitely be
greater than $k$, since $\transtheo$ contains the constraint
$\transl{\varphi}$. However, since $J_i$ is completely unknown on
dynamic predicates, we see that  $J_i\leqp J_0$. 
Hence, using Proposition \ref{prop:kleeneprecise}, we find that also $J_0\not\models\transl{\varphi}$, which is in contradiction with the assumption that $\pss{\struct}{0}\models  \inittheo$.
 For bistate sentences, a similar argument holds. 
 Thus we can conclude that indeed for every sentence $\varphi$ in \theory, $\kleeneval_J(\varphi)\neq\lfalse$.
 
Now, let \D be the definition of \theory. We should show that $\kleeneval_J(\D) \neq\lfalse$.  As \D is stratified over time, by Theorem 4.5 in \cite{tocl/VennekensGD06}, we can split \D in definitions $(\D_i)_{i\in\nat}$ for each time point. 
The definitions $\D_i$ with $i\leq k$ are satisfied in $\struct$ because those are the definitions in the theory \transtheo. 
For definitions $\D_i$ with $i>k$, these definitions define only dynamic atoms with   
dynamic arguments greater than $k$. 
Furthermore, these dynamic atoms are completely unknown in \struct. 
Proposition \ref{prop:defunkn} then yields that $\kleeneval_\struct(\D)=\lunkn\neq\lfalse$.

Thus, we also find that $\kleeneval_\struct(\theory)=\lunkn\neq\lfalse$, i.e.,  \struct is  indeed weakly \theory-compatible.
%  \maurice{Ik ben niet mee, moet je niet aantonen dat er een
%    contradictie als voor $i<k$ $J_i$ een model is voor $\varphi$ en
%    $J_{i+1}$ geen model is?}
% BART: neen, voor sentences bewijs ik het uit het ongerijmde, voor definties niet... Duidelijker nu???
 \end{proof}

\noindent
\emph{Corollary~3.20}

\noindent
 LTC-theories satisfy the Markov property and the weak Markov property.

\begin{proof}
 We first prove that LTC theories satisfy the Markov property.
 Let $J$ and $J'$ be a $k$-chain and a $k'$-chain respectively with $J_k = J'_{k'}$. 
 Suppose $S$ is a \theory-successor of $J'$. We show that $S$ is also a \theory-successor of $J$. 
 Since $\extension{J'}{S}$ is \theory-compatible, there exists a model $K'$ of \theory such that $K_i'=J'_i$ for $i\leq k'$ and $K_{k'+1}=S$.
 Now let $K$ be the structure such that
 \[K_j=\left\{\begin{array}{ll}
          J_j & \text{for } j\leq k,\\
K'_{k'+j-k} &\text{otherwise.}
  \end{array}
  \right.
 \]
We claim that $K$ is a model of \theory more precise than $\extension{J}{S}$. The fact that it is more precise than $\extension{J}{S}$ follows from the fact that $K_{k+1}=K'_{k'+(k+1)-k}=K'_{k'+1}$, which equals $S$, by construction of $K$. 
 In order to prove our claim, we show that for every $j$, $(K_j,K_{j+1})$, satisfies \transtheo.
 For $j\leq k$, this follows from the fact that $J$ is \theory-compatible; for $j>k$, from the fact that $K'$ is a model of \theory.
 Now using Theorem 3.18, we see that $K$ is a model of \theory, which shows that $\extension{J}{S}$ is indeed \theory-compatible.
 
 We now prove that LTC theories satisfy the weak Markov property.
 This follows immediately from Theorem 3.19: \extension{J}{S} is weakly \theory-compatible if and only if $J$ is weakly \theory-compatible and  $(J_k,S)\models\transtheo$.
\end{proof}

\noindent
\emph{Theorem 4.2}

\noindent
 Let \theory be an LTC-theory and $\varphi$ a universal single-state sentence. 
 Then $\theory\models\varphi$ if $\inittheo \models \transl{\varphi}$,
 and $(\transtheo \land \transl{\varphi} ) \models
 \transl{\varphi[\next(t)/t]}$, where $t$ is the unique time-variable
 in $\varphi$.

\begin{proof}%[Proof of Theorem 4.2]
 This theorem is in fact a reformulation of the principle of proofs by induction. The condition $\inittheo \models \transl{\varphi}$ expresses that the invariants holds at time $0$, i.e., this is the base case. The condition $(\transtheo \land \transl{\varphi} ) \models \transl{\varphi[\next(t)/t]}$ expresses that whenever the invariants holds at $t$, it also holds at $\next(t)$.
\end{proof}

\noindent
\emph{Theorem 4.3}

\noindent
	 Let $\struct$ be a $\Vs{\voc}$-structure and let $\varphi$ be a universal single-state sentence with time variable $t$.
	 Then $\varphi$ is satisfied in all \voc-structures expanding $\struct$ if $\inittheo \land \lnot \transl{\varphi}$ has no models expanding $\struct$, and 
		 $\transtheo \land \transl{\varphi}  \land \lnot  \transl{\varphi[\next(t)/t]}$  has no models expanding $\struct$.

\begin{proof}%[Proof of Theorem 4.3]
 This theorem is also a reformulation of the principle of proofs by induction, analogue to Theorem 4.2. 
\end{proof}

\noindent
\emph{Theorem 4.4}

\noindent
 Let \theory be an LTC-theory and $\varphi$ a universal bistate sentence. 
 Then $\theory\models\varphi$ if and only if $\transtheo \models \transl{\varphi}$.

 \begin{proof}
%  One direction, is clear: if $\theory\models\varphi$, it follows immediately that  $\transtheo \models \transl{\varphi}$.
  
  Suppose $\transtheo \models \transl{\varphi}$. We should show that $\theory\models\varphi$. Therefore, let $J$ be a model of $\theory$. By Theorem 3.18, for every $k$, $J_k\models \transtheo$. Thus, using our assumption, for every $k$, also $J_k\models  \transl{\varphi}$. 
  But $\varphi$ is itself an LTC-theory, and $\varphi_i=\ltrue$ and $\varphi_t = \transl{\varphi}$. Thus, using Theorem 3.18 again, we find that $J\models \varphi$, as desired.
 \end{proof}

\end{document}